\documentclass[twocolumn]{aastex631}
\usepackage{amsmath} 

\begin{document}

\title{A shock crashing into confined dense circumstellar matter brightens the nascent SN~2023ixf }

\author[0000-0003-3031-6105]{Maokai Hu}
\affiliation{Physics Department, Tsinghua University, Beijing 100084, China}
\email{kaihukaihu123@mail.tsinghua.edu.cn}
\author[0000-0001-7092-9374]{Lifan Wang}
\affiliation{George P. and Cynthia Woods Mitchell Institute for Fundamental Physics \& Astronomy, Texas A. \& M. University, 4242 TAMU, College Station, TX 77843, USA}
\email{lifan@tamu.edu}
\author[0000-0002-7334-2357]{Xiaofeng Wang}
\affiliation{Physics Department, Tsinghua University, Beijing 100084, China} 





\begin{abstract}

Red supergiants may experience a short-lived period of episodic mass loss rather than steady winds before their core collapses, leading to dense circumstellar matter (CSM) close to core-collapse supernovae (SNe). Interaction of SN ejecta with such nearby CSM can generate additional radiation, appending to the cooling radiation from the shock breakout of the progenitor envelope, to brighten the nascent SN explosion. This phenomenon is conspicuous for SN~2023ixf as its $V$-band brightness showed a rapid increase of about three magnitudes from the first to the third day after the time of the first light, which is distinctive among type II SNe with flash ionized signatures. In this paper, we employed a Monte Carlo method to simulate the radiative diffusion process in the unshocked CSM. Considering a wide range of mass-loss rates from $10^{-5}$ to $10^{-2}$ M$_{\odot}$\,yr$^{-1}$, we confirmed that the fast-rising light curve of SN~2023ixf can be fitted by the interaction of the SN ejecta with a CSM having a mass-loss rate of about $10^{-2}$ M$_{\odot}$\,yr$^{-1}$ located within $10^{15}$ cm to the progenitor. 

\end{abstract}



\section{Introduction} \label{sec:intro}

When massive stars deplete their nuclear fuel, the collapse of the core creates an explosive event known as core-collapse supernovae (SNe). Several pre-explosion detections support the scenario that red supergiants (RSGs) are the progenitor of type II SNe with plateau-shaped light curves (e.g., \citealt{2009ARA&A..47...63S}). Thanks to the wide-field optical survey programs and the fast-responding spectroscopically follow-up observations, a relatively large fraction of type II SNe were found to display short-lived narrow emission lines in their early-phase spectra \citep{2014Natur.509..471G,2016ApJ...818....3K,2017NatPh..13..510Y,2018ApJ...861...63H,2020MNRAS.498...84Z,2021ApJ...912...46B,2021ApJ...907...52T}. These emission lines arise from the recombination of the highly ionized gas in the compact circumstellar matter (CSM) due to the emerging shock breakout \citep{2019MNRAS.483.3762K}. Apart from the cooling radiation of the shock breakout, the subsequent interaction between the ejecta and the nearby CSM could also generate additional radiation, resulting in an early-phase fast-rising light curve relating to a scenario of an RSG embedded in the dense confined CSM.

SN~2023ixf, a type II SN exploded in Messier 101 (M101) \citep{2023TNSTR1158....1I} and was serendipitously captured by amateur astronomers; this provides a crucial chance for constraining the SN progenitor systems.
The intra-night observations of SN~2023ixf over the first few days since the discovery delineate the early-phase evolution of its ionized flash signatures. The narrow emission lines from the recombination of the ionized gas rapidly decrease and mostly disappear a week later, suggesting the existence of dense circumstellar matter (CSM) close to the progenitor star \citep{2023ApJ...956L...5B,2023ApJ...955L...8H,2023ApJ...954L..42J,2023ApJ...956...46S,2023arXiv230901998Z}. Pre-explosion archival images from the HST and the Spitzer Space Telescope indicate that the progenitor of SN~2023ixf is an RSG with a dusty environment and it experienced eruptive or episodic mass loss shortly before the explosion \citep{2023ApJ...957...28D,2023ApJ...952L..30J,2023ApJ...952L..23K,2023ApJ...957...64S,2024MNRAS.527.5366N,2024SCPMA..6719514X}. Therefore, SN~2023ixf is an excellent example to investigate the mass-loss history of massive stars before core collapse. 

M101 is an attractive target to amateur astronomers who monitor it frequently, leading to the first detection of SN~2023ixf about one day earlier than the discovery \citep{2023TNSAN.130....1M}. These rare chromatic images recorded the radiation from the shock breakout of the RSG envelope and exhibited a color redder than the typical model predictions. \cite{2023arXiv231114409L} suggest that there is time-variant extinction due to the sublimation of circumstellar dust during the spherical phase of the shock breakout. On the other hand, shock cooling and the ejecta$-$CSM interaction contribute to the radiation a few days after the explosion, resulting in the fast-rising light curve of SN~2023ixf \citep{2023arXiv231114409L,2023arXiv231010727Z}. 

In this paper, we focus on the ejecta$-$CSM interaction process to fit the fast-rising light curve of SN~2023ixf. Section~\ref{SecII} describes our Monte Carlo method to simulate the radiative diffusion process in the unshocked CSM, and Section~\ref{SecIII} displays the results. The discussions and conclusions are given in Section~\ref{SecIV}. 

\section{Methods} 
\label{SecII}

The highlight finding in \cite{2023arXiv231114409L} is the exotic color evolution of SN~2023ixf during its extremely early phase, that the $g-r$ color rapidly decreases from about $0.8$ to $0.0$ within the interval between +1.41 and +5.66 hours after the time of the first light. This red-to-blue evolution is contradictory to the shock breakout theory since the cooling temperature from the shock breakout of an RSG decreases from $\sim 10^{5}$ K to $\sim 10^{4}$ K within a few hours since the explosion (e.g., \citealt{2010ApJ...725..904N}). A natural possibility of the red-to-blue evolution is the time-variant extinction relating to the gradually destroyed circumstellar dust as the progenitor star of SN~2023ixf is embedded in dense CSM. The survival time of circumstellar dust may be prolonged to a few hours by considering an asymmetric shock breakout, the diffusion process within the ionized CSM, and the dust size. 

We followed the hybrid model adopted in \cite{2023arXiv231114409L} to incorporate the radiation from the cooling radiation of shock breakout and the radiation from CSM interaction to match the early-time light curve of SN~2023ixf. The additional aspect of our work involves considering the radiative diffusion process within the unshocked CSM for the photons generated either from shock cooling or CSM interaction. Firstly, we briefly describe the cooling radiation from the shock breakout of the RSG envelope and the radiation from the CSM interaction. Then, we introduce our Monte Carlo method to simulate the diffusion process within the unshocked CSM.  

\subsection{Cooling radiation from the shock breakout of an RSG envelope}

A shock breakout emerges when the shock reaches the edge of the star, in which the optical depth is approximate $c/V_{\rm sh}$, where $V_{\rm sh}$ is the shock velocity, and $c$ is the light speed. Meanwhile, the subsequent expanding cooling envelope can contribute to the early-time emission on a timescale of about a few days. In the case of the shock breakout of an RSG, the corresponding emission is in thermal equilibrium. The luminosity ($L$) and temperature ($T$) evolve as below \citep{2010ApJ...725..904N},
\begin{equation}
\label{eqN_1}
\begin{aligned}
L = & 10^{44}\ {\rm erg}\,{\rm s}^{-1}\ M_{15}^{-0.37}R_{500}^{2.46}E_{51}^{0.3}t_{\rm hr}^{-4/3} \\  
T = & 10\ {\rm eV}\ M_{15}^{-0.22}R_{500}^{0.12}E_{51}^{0.23}t_{\rm hr}^{-0.36}
\end{aligned}
\end{equation}
where $M_{15}$, $R_{500}$, and $E_{51}$ are the progenitor mass ($M_{\rm star}$), progenitor radius ($R_{\rm star}$), and the explosion energy ($E_{\rm ej}$) in units of 15 ${\rm M}_{\odot}$, 500 ${\rm R}_{\odot}$, and $10^{51}\ {\rm erg}$, respectively. $t_{\rm hr}$ is the time since the explosion with the hour unit. The above equation is applicable to  $t < t_{\rm tr}$, where $t_{\rm tr}$ is a transition time from the planar phase to spherical phase expressed as $t_{\rm tr} = 14\ {\rm hr}\ M_{15}^{0.43}R_{500}^{1.26}E_{51}^{-0.56}$ \citep{2010ApJ...725..904N}. For $t > t_{\rm s}$, the corresponding luminosity and temperature are expressed as \citep{2010ApJ...725..904N},
\begin{equation}
\label{eqN_2}
\begin{aligned}
L =  & 3\times10^{42}\ {\rm erg}\,{\rm s}^{-1}\ M_{15}^{-0.87}R_{500}^{1.0}E_{51}^{0.96}t_{\rm d}^{-0.17} \\ 
T = & 3\ {\rm eV}\ M_{15}^{-0.13}R_{500}^{0.38}E_{51}^{0.11}t_{\rm d}^{-0.56}
\end{aligned}
\end{equation}
where $t_{\rm d}$ is the time with the unit of day. 
The shock cooling emission occurs below the unshocked CSM, and the simplification of ignoring the diffusion effect within the unshocked CSM may introduce some bias. In this paper, the photons calculated from Equation~\ref{eqN_1} and Equation~\ref{eqN_2} are injected into the unshocked CSM to generate the final light curve of the shock cooling component through our Monte Carlo procedure. 

\begin{figure}[t!]
\centering
\includegraphics[width = 0.45 \textwidth]{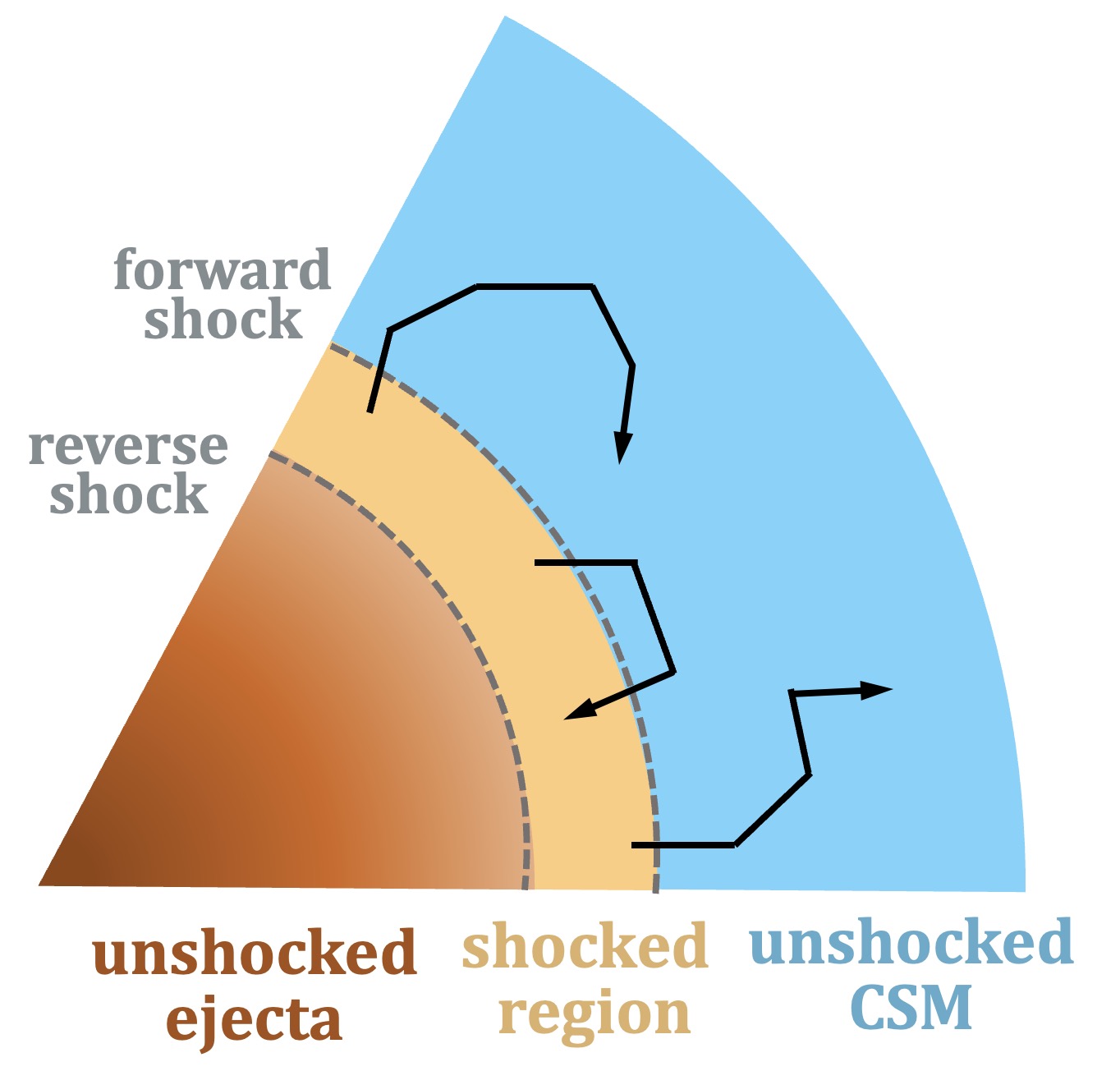}
\caption{The illustration of the ejecta$-$CSM interaction process. The tangerine, yellow, and blue regions are separated by the forward shock and reverse shock (two dashed gray lines) and represent the unshocked ejecta, the shocked region, and the unshocked CSM, respectively. The photons generated from CSM interaction are injected into the unshocked CSM, as shown by the short black lines.} 
\label{fig_11} 
\end{figure}

\subsection{Ejecta-CSM interaction}

Ejecta$-$CSM interaction is an efficient engine to convert kinetic energy into radiation and power the light curve of SNe with mass-loss history prior to the explosion \citep{1994ApJ...420..268C,Wood-Vasey:2004ApJ...616..339W,2012ApJ...759..108S,2013MNRAS.435.1520M,2020PASJ...72...67T}. In particular, the presence of confined CSM is responsible for the fast-rising light curve of core-collapse SNe \citep{2018NatAs...2..808F} and the early excess observed in thermonuclear SNe \citep{2021ApJ...923L...8J,2023MNRAS.525..246H,2023MNRAS.521.1897M,2023MNRAS.522.6035M}.

Analytic solutions of the eject$-$CSM interaction have been studied for situations where the unshocked CSM is optically thick or thin \citep{1982ApJ...258..790C,1994ApJ...420..268C,2011MNRAS.415..199M,2013MNRAS.435.1520M}. In the case of optically thin CSM (e.g., the corresponding optical depth of unshocked CSM, $\tau_{\rm csm} < 1.0$), the radiation from the CSM interaction can directly escape from the unshocked CSM. When the unshocked CSM is optically thick (e.g., $\tau_{\rm csm} \ge 30$), the photons may be trapped behind the shock, leading to a shock breakout as the shock emerges from the CSM. However, for the case where the unshocked CSM is moderately optically thick (e.g., $\tau_{\rm csm} \sim 10$), the time-scale of photons escaping from the unshocked CSM is shorter than that of the shock propagating in CSM, but the photons will undergo a diffusion process within the unshocked CSM relating to the Thomson scattering. A Monte Carlo program is practicable to simulate the photon diffusion process within the unshocked CSM during the interaction. 

In this paper, we adopt the models of \cite{2013MNRAS.435.1520M} and \cite{2023MNRAS.525..246H} to describe the CSM interaction process and update the Monte Carlo method in \cite{2022ApJ...931..110H} to simulate the radiative diffusion process in the unshocked CSM. A schematic illustration of the CSM interaction and the photon diffusing process is shown in Figure~\ref{fig_11}, in which the scheme is divided into three regions as the unshocked CSM, the shocked region, and the unshocked ejecta. The thickness of the shocked region is identified by the forward and reverse shock calculated from \cite{1982ApJ...258..790C}.

\begin{figure*}[!t]
\centering
\includegraphics[width = 0.85 \textwidth]{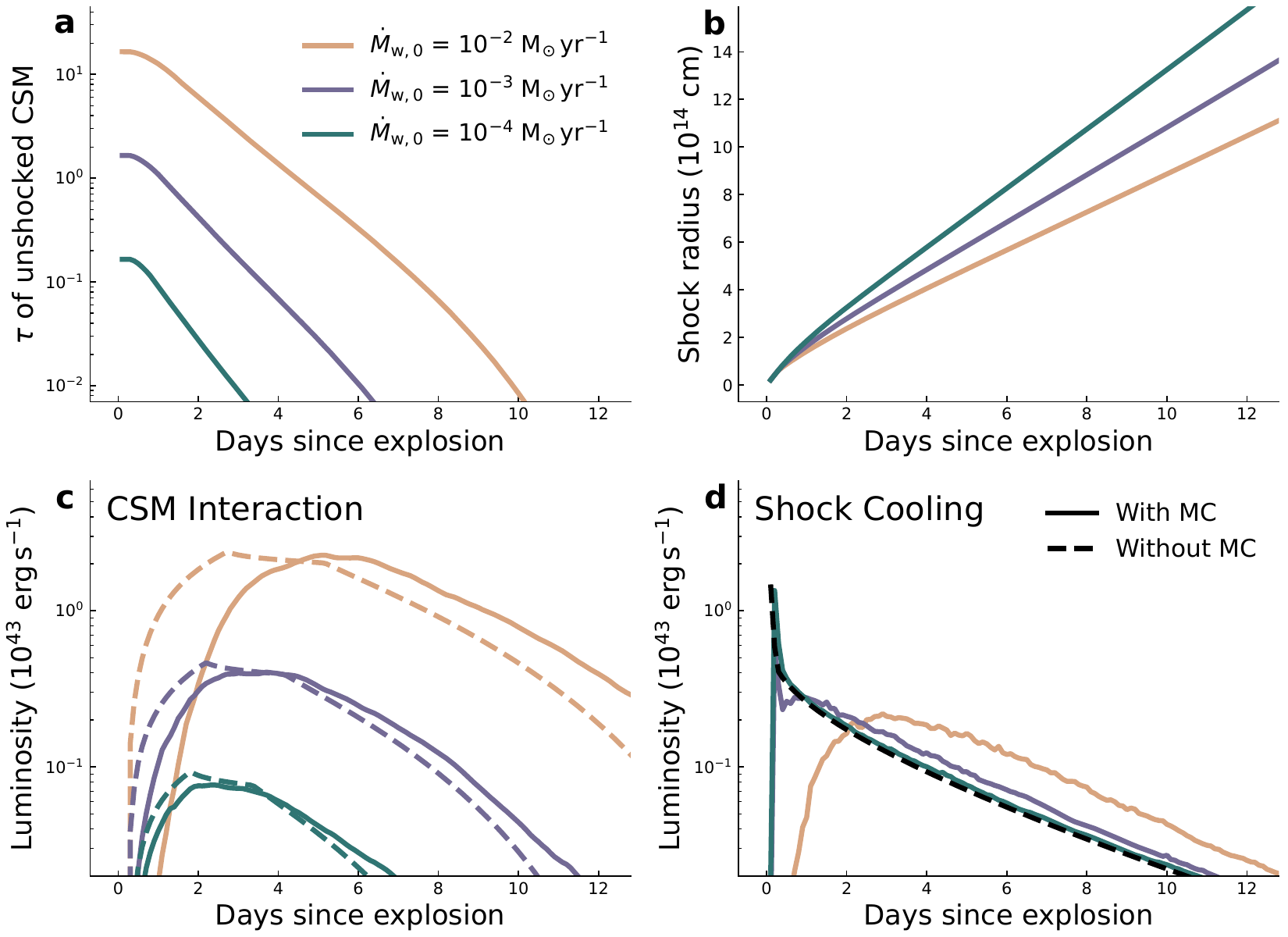}
\caption{The performances of our Monte Carlo procedure with three cases of $\dot{M}_{\rm w,0} = 10^{-2}\ {\rm M}_{\odot}\,{\rm yr}^{-1}$ (yellow lines), $10^{-3}\ {\rm M}_{\odot}\,{\rm yr}^{-1}$ (purple lines), and $10^{-4}\ {\rm M}_{\odot}\,{\rm yr}^{-1}$ (green lines), respectively. {\bf a}: the radial optical depth of unshocked CSM ($\tau_{\rm csm}$) with the three cases of $\dot{M}_{\rm w,0}$, indicating that the photon diffusing process should be considered with $\dot{M}_{\rm w,0}$ of around $10^{-2}\ {\rm M}_{\odot}\,{\rm yr}^{-1}$. {\bf b}: the shock radii related to the CSM interaction. {\bf c}: comparisons of the bolometric luminosity of the CSM interaction with/without the Monte Carlo (MC) process. The cases with the Monte Carlo process (solid lines) simulate the diffusion process within the unshocked CSM, while the cases without the Monte Carlo process (dashed lines) assume that the emission could directly escape from the unshocked CSM. {\bf d}: same as the panel c but for the bolometric luminosity of the shock cooling. The dashed black line is the shock cooling luminosity calculated from Equation~\ref{eqN_1} and Equation~\ref{eqN_2} without considering the diffusion effect.} 
\label{fig_22} 
\end{figure*} 


\begin{figure*}[t]
\centering
\includegraphics[width = 0.85\textwidth]{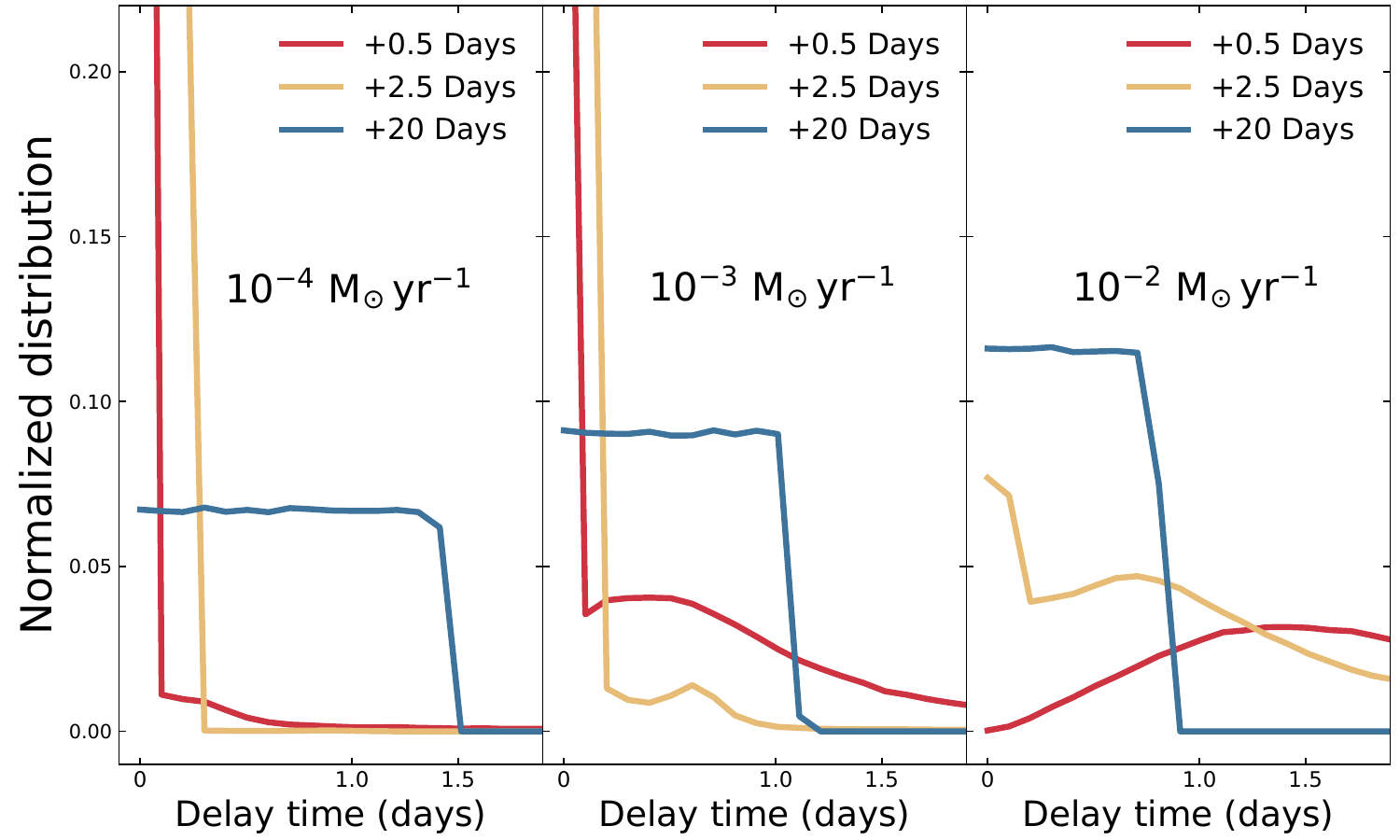}
\caption{The red, yellow, and blue lines are the normalized kernel distribution for a radiation pulse injected into the unshocked CSM at +0.5 days, +2.5 days, and +20 days since the explosion, respectively. The left, middle, and right panels relate to the $\dot{M}_{\rm w,0}$ of $10^{-4}\ {\rm M}_{\odot}\,{\rm yr}^{-1}$, $10^{-3}\ {\rm M}_{\odot}\,{\rm yr}^{-1}$, and $10^{-2}\ {\rm M}_{\odot}\,{\rm yr}^{-1}$, respectively. The delay time is due to the propagation effect of the generated photons associated with the CSM interaction, which involves two factors: the diffusion process within the unshocked CSM and the shell-like radiative region.}
\label{fig_kernel}
\end{figure*}

As shown in \cite{2025ApJ...978L..27H}, we adopt a similar formula to describe the radial density profile of the CSM, 
\begin{scriptsize}
\begin{align} 
\label{eq_Mw}
\dot{M}_{\text{w}}(R) = \begin{cases} 
0, & R \le R_0 \\
\dot{M}_{\rm w,0}(\frac{R - R_0}{R_1 - R_0})^{n_1}, & R_0 < R \le R_1 \\ 
\dot{M}_{\rm w,0}, & R_1 < R \le R_2 \\
(\dot{M}_{\rm w,0} - \dot{M}_{\rm w,min}) (\frac{R_3 - R}{R_3 - R_2})^{n_2} + \dot{M}_{\rm w,min}, & R_2 < R \le R_3 \\
\dot{M}_{\rm w,min}, & R > R_3\\
\end{cases}
\end{align} 
\end{scriptsize}
where $R$ is the distance to the SN, and $\dot{M}_{\rm w}$ is the mass-loss rate of the CSM, and  $n_1$ and $n_2$ are two parameters depicting how fast $\dot{M}_{\rm w}(R)$ increases from 0 to $\dot{M}_{\rm w,0}$ and decreases from $\dot{M}_{\rm w,0}$ to $\dot{M}_{\rm w,min}$, respectively. The distance-variant distribution of $\dot{M}_{\rm w}(R)$ between the radii $R_0$ and $R_3$ may relate to the eruptive mass-loss history of the progenitor, and the constant mass-loss rate ($= \dot{M}_{\rm w,min}$) corresponds to the usual stellar wind scenario. This hybrid scenario is possible in massive stars with local super-Eddington luminosity in stellar envelopes \citep{2024arXiv240512274C}.
The CSM density follows the expression as $\rho_{\rm csm} = \dot{M}_{\rm w}/(4\pi R^2v_{\rm w})$, where $v_{\rm w}$ is the wind speed and is set to 75 km\,s$^{-1}$ consistent with the previous studies in SN~2023ixf. 

For the homogeneous expansion, the density of exterior ejecta satisfies the power-law profile as $\rho_{\rm ej} \propto R^{-n}$, where $n = 12$ corresponding to the explosion of an RSG. The dynamic procedure of the ejecta crashing into CSM can be solved by the equation below, 
\begin{equation} 
\label{eq1} 
M_{\text{sh}}\frac{\mathrm{d}V_{\text{sh}}}{\mathrm{d}t} = 4\pi R_{\text{sh}}^2[\rho_{\text{ej}}(v_{\text{ej}} - V_{\text{sh}})^2 - \rho_{\text{csm}}(V_{\text{sh}} - v_{\rm w})^2]
\end{equation} 
where $M_{\rm sh}$, $R_{\rm sh}$, and $V_{\rm sh}$ are the total mass, radius, and shock velocity, respectively. Thus, the bolometric luminosity is given as $L = \frac{\epsilon}{2}\frac{\dot{M}_{\text{w}}}{v_{\rm w}}V_{\text{sh}}^3$, where $\epsilon$ is a conversion efficiency with the value of 0.6 in our study. This moderately large value of $\epsilon$ is related to the moderately large mass-loss rate considered in the case of SN~2023ixf. 

The luminosity and black-body temperature can be numerically solved based on Equation~\ref{eq_Mw} and \ref{eq1}, and we can inject photons into the shocked region and unshocked CSM. Then, we can track the photon diffusing process with the Monte Carlo method to generate the final light curve. 

\subsection{The Monte Carlo method}

\begin{figure*}[t!]
\centering
\includegraphics[width = 0.9 \textwidth]{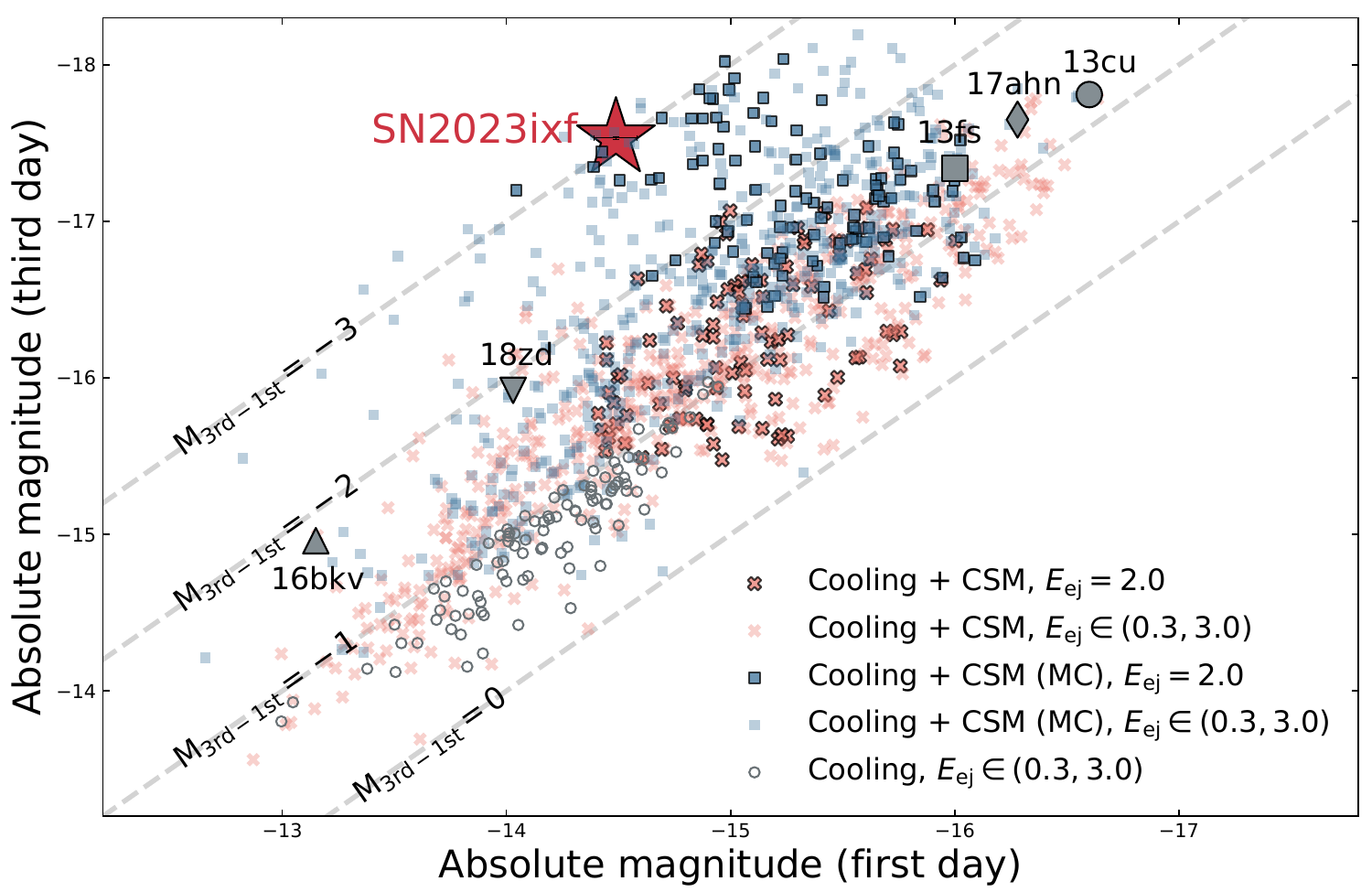}
\caption{The absolute magnitude of the first day versus the third day after the time of the first light for SN~2023ixf (red pentacle, $V$-band, \citealt{2023arXiv231114409L}), SN~2013cu (gray circle, $r$-band, \citealt{2014Natur.509..471G}), SN~2013fs (gray square, $r$-band, \citealt{2017NatPh..13..510Y}), SN~2016bkv (gray up-triangle, clear-band, \citealt{2018ApJ...861...63H}), SN~2017ahn (gray diamond, $V$-band, \citealt{2021ApJ...907...52T}), and SN~2018zd (gray down-triangle, clear band, \citealt{2020MNRAS.498...84Z}). The pink cross symbols are the predicted distribution without considering the diffusion process and with the range of $\dot{M}_{\rm w,0}$ from $10^{-5}$ to $10^{-3}\ {\rm M}_{\odot}\,{\rm yr}^{-1}$, and the cyan-blue square symbols correspond to the cases with Monte Carlo process to simulate the diffusion process and with the range of $\dot{M}_{\rm w,0}$ from $10^{-4}$ to $10^{-2}\ {\rm M}_{\odot}\,{\rm yr}^{-1}$. The pink and cyan-blue symbols with black edges represent that the kinetic energy of the ejecta is set to $2.0\times10^{51}\ {\rm erg}$, which is consistent with the result in previous studies of SN~2023ixf. For the comparison, the predicted distribution of absolution magnitudes relating to the pure shock breakout of the envelope without CSM interaction is shown with hollow gray circles. The dashed gray lines represent the magnitude difference between the third and first days since the first light ($M_{\rm 3rd-1st}$) of $M_{\rm 3rd-1st} = -3$, $M_{\rm 3rd-1st} = -2$, $M_{\rm 3rd-1st} = -1$, $M_{\rm 3rd-1st} = 0$, respectively.} 
\label{fig_33} 
\end{figure*}

The Monte Carlo process is a well-established method for solving the radiative transfer problem, such as the light echoes of dusty CSM and the line formation of SN ejecta (e.g., \citealt{Witt1977ApJS...35....1W,1999A&A...345..211L,2001ApJ...551..269G,2006ApJ...651..366K,2013ARA&A..51...63S,2014ApJ...780...18G,2017ApJ...834..118M,Ding2021,2022ApJ...931..110H,2023ApJ...953..132W}). In this study, we have employed the basic configuration of the Monte Carlo process developed in \cite{2022ApJ...931..110H}, which includes launching photons into CSM, scattering and absorption due to circumstellar dust, and collecting the photons escaping from the CSM. However, several differences exist between the Monte Carlo process in this paper and \cite{2022ApJ...931..110H}: 

a) The photon-CSM interaction in this study is Thomson scattering due to the ionized gas rather than the Mie scattering from the circumstellar dust. We assume the opacity relating to the Thomson scattering is 0.32 cm$^2\,$g$^{-1}$, and we ignore the opacity due to free-free, bound-free, and bound-bound absorption.

b) The physical state (e.g., the optical depth and the boundary of the circumstellar dust in \cite{2022ApJ...931..110H}) is time invariable, as the circumstellar dust is located at the distance of $10^{17}$ cm to the SN. At such a location, the circumstellar dust cannot be destroyed for a few years until the ejecta collides with the dust. In this paper, the characteristic distance of CSM to the SN is less than $10^{15}$ cm, producing a rapid dynamic process of the ejecta$-$CSM interaction. In this scenario, the physical state evolves rapidly, such as the optical depth and the inner boundary of the unshocked CSM (approximately equal to the shock radius) shown in panels a and b of Figure~\ref{fig_22}, making the Monte Carlo simulation complicated and time-consuming. 

c) The photons injected into the circumstellar dust could be regarded as from a point-like source in \cite{2022ApJ...931..110H}, since the SN photosphere is much smaller than the distance of circumstellar dust to the SN. However, the geometric size of the emitting region can not be ignored in this paper, which increases from a point-like source to a shell-like source comparable to the scale of the unshocked CSM within a few days since the explosion. A radiative pulse from a shell-like source would have a plateau-like distribution from a distant observer. On the other hand, for the photon scattered into the unshocked ejecta, we assume a probability ($\sim 0.9$) that the photon is reflected at the inner boundary of the shocked region (the position of the reverse shock as shown in Figure~\ref{fig_11}) for the subsequent Monte Carlo process. The reflection process is similar to the method shown in \cite{2014ApJ...780...18G}. This simplification makes us focus on the photon diffusing process within the unshocked CSM and avoid the arduous radiative transfer process in the unshocked ejecta due to the complicated distribution of its density and compositions. 

Figure~\ref{fig_22} shows the performances of our Monte Carlo process with three cases of $\dot{M}_{\rm w,0} = 10^{-2}\ {\rm M}_{\odot}\,{\rm yr}^{-1}$, $10^{-3}\ {\rm M}_{\odot}\,{\rm yr}^{-1}$, and $10^{-4}\ {\rm M}_{\odot}\,{\rm yr}^{-1}$, respectively. For each case, they share the same parameter setting as $R_0 = 5\times10^{13}\ {\rm cm}$, $R_1 = 3\times10^{14}\ {\rm cm}$, $R_2 = 5\times10^{14}\ {\rm cm}$, $R_3 = 1.5\times10^{15}\ {\rm cm}$, $R_{\rm star} = 350\ {\rm R}_{\odot}$, $M_{\rm star} = 15\ {\rm M}_{\odot}$, and $E_{\rm ej} = 2.0\times10^{51}\ {\rm erg}\,{\rm s}^{-1}$, which are consistent with the result of fitting the early-time light curve of SN~2023ixf in this paper. The panel~a of Figure~\ref{fig_22} displays the relationship between the optical depth of the unshocked CSM and the parameter $\dot{M}_{\rm w,0}$, indicating that it is necessary to consider the photon diffusion process within the unshocked CSM with the $\dot{M}_{\rm w,0}$ of about $10^{-3} - 10^{-2}\ {\rm M}_{\odot}\,{\rm yr}^{-1}$. The diffusion effect could reshape the luminosity curve for the relatively high mass-loss rate of CSM as shown in panels c and d of Figure~\ref{fig_22}. Another consequence is the interval between the explosion time and the time of the first light because the photons would be trapped within the unshocked CSM for minutes or hours, depending on the geometric distribution of CSM. 


\begin{figure*}[t!]
\centering
\includegraphics[width = 0.85 \textwidth]{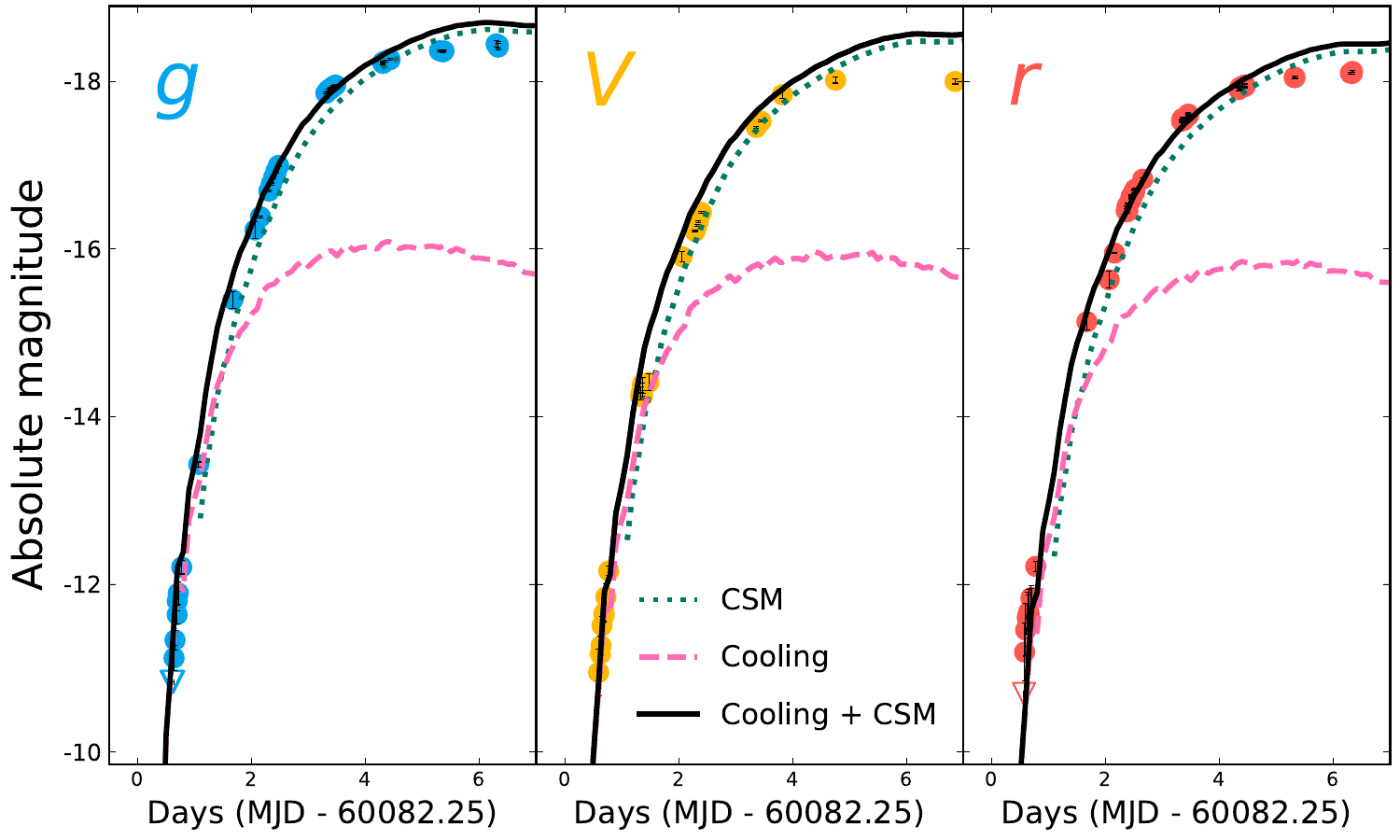}
\caption{The result of our CSM model with Monte Carlo process to fit the early-time light curve of SN~2023ixf. The blue, yellow, and red circles represent the photometric data of $g$, $V$, and $r$ bands, respectively, from \cite{2023arXiv231114409L}. For each panel, the dashed pink, dotted green, and solid black lines represent the components of shock cooling, CSM interaction, and their combination.} 
\label{fig_44} 
\end{figure*}

To clearly illustrate the effect of the diffusion process, Figure~\ref{fig_kernel} displays the behavior of a radiative pulse injected into the unshocked CSM concerning different epochs and mass-loss rates. The parameter setting is the same as in Figure~\ref{fig_22}. For the case of $\dot{M}_{\rm w,0} = 10^{-4}\ {\rm M}_{\odot}\,{\rm yr}^{-1}$, the unshocked CSM is approximately optically thin, and the generated radiation can directly escape from the unshocked CSM. Therefore, the observed kernel distribution at +0.5 days after the explosion also has a pulse-like shape, and the radiative region relating to the CSM interaction could be regarded as a point-like source. As the shock propagates, the shock radius increases to $\sim 10^{15}$ cm (as shown in the panel b of Figure~\ref{fig_22}), resulting in a plateau-like shape of the kernel distribution due to the shell-like radiative region instead of a point-like source. On the other hand, the case of $\dot{M}_{\rm w,0} = 10^{-2}\ {\rm M}_{\odot}\,{\rm yr}^{-1}$ has a different kernel distribution at +0.5 days after the explosion, as the existence of the radiative diffusion process within the unshocked CSM.


\section{Results} 
\label{SecIII}

The pre-explosion images suggest that SN~2023ixf has a progenitor relating to an RSG with a significant mass-loss history, which is also confirmed by the emergence of its first light. Here, we mainly concentrate on the absolute magnitude on the first day ($M_{\rm 1st}$) and the third day ($M_{\rm 3rd}$) after the time of the first light and adopt their magnitude difference ($M_{\rm 3rd - 1st}$) as an indicator to probe the brightening of the early-time light curve. This simplified parameter can be directly obtained from the observational data and the predicted light curve by the model. Meanwhile, it can also reasonably reduce the influence of the radiation from the recombination of ionized hydrogen near the peak brightness.

In Figure~\ref{fig_33}, we can see that the $M_{\rm 3rd - 1st}$ of SN~2023ixf in the $V$ band is about $-3.0$, which is significantly higher than other SNe II with flash spectroscopic features. Note that the $M_{\rm 3rd - 1st}$ of other SNe includes $V$, $r$, and clear bands due to the challenge of catching early-phase signals and is based on the explosion time from their references. Although the filter curves in these three optical bands and the interval between the explosion time and the first light may introduce some minor bias in the $M_{\rm 3rd - 1st}$, the comparison still indicates that SN~2023ixf may differ in terms of explosive energy, the radial distribution of CSM, and even the asymmetric distribution of CSM. 

To uncover the physical nature behind SN~2023ixf, we randomly generate hundreds of parameter sets to predict the corresponding $M_{\rm 3rd - 1st}$ of each model. The radius and mass of progenitors vary from 100 to 900 ${\rm R}_{\odot}$ and from 10 to 20 ${\rm M}_{\odot}$, respectively. The ejecta kinetic energy ranges from 0.3 to $3.0\times10^{51}\ {\rm erg}$. The $\dot{M}_{\rm w,0}$ varies from $10^{-5}$ to $10^{-3}$ ${\rm M}_{\odot}\,{\rm yr}^{-1}$ for CSM interaction model without considering the diffusion effect, and from $10^{-4}$ to $10^{-2}$ ${\rm M}_{\odot}\,{\rm yr}^{-1}$ for cases with the Monte Carlo process. The CSM distance to SN for both situations ranges from $10^{14}$ to $10^{15}$ cm. The pink cross and cyan-blue square symbols in Figure~\ref{fig_33} display the predicted $M_{\rm 3rd - 1st}$ relating to each randomly generated parameter set. Note that the explosion time is just the time of the first light for the cases without considering the diffusion process, while we estimate the time of the first light as the time when the absolute magnitude equals to -10.0 for the cases with the Monte Carlo process. 

It clearly shows that the observed $M_{\rm 3rd - 1st}$ of SNe~2013cu, 2013fs, 2016bkv, 2017ahn, and 2018zd may be well-matched with the model for both considering the diffusion process or not, and the observed $M_{\rm 3rd - 1st}$ of SN~2023ixf can only be fitted by the CSM model with the Monte Carlo process, indicating the relatively high-density CSM around SN~2023ixf. In particular, we adopt the similar progenitor radius ($R_{\rm star} = 350\ {\rm R}_{\odot}$), progenitor mass ($M_{\rm star} = 15\ {\rm M}_{\odot}$), and the explosion energy ($E_{\rm ej} = 2.0\times10^{51}\ {\rm erg}\,{\rm s}^{-1}$) as in previous studies, but with the randomly generated $\dot{M}_{\rm w,0}$ and CSM distance to SN, to predict the corresponding $M_{\rm 3rd - 1st}$ shown in Figure~\ref{fig_33} as the pink or cyan-blue symbols with black edges, supporting the conclusion in previous studies.

Moreover, we find the best-fit model for the early-phase light curve of SN~2023ixf shown in Figure~\ref{fig_44}, which has corresponding CSM parameters as $R_0 = 5\times10^{13}\ {\rm cm}$, $R_1 = 3\times10^{14}\ {\rm cm}$, $R_2 = 5\times10^{14}\ {\rm cm}$, $R_3 = 1.5\times10^{15}\ {\rm cm}$, $\dot{M}_{\rm w,0} = 0.9\times10^{-2}\ {\rm M}_{\odot}\,{\rm yr}^{-1}$, and $E_{\rm ej} = 1.9\times10^{51}\ {\rm erg}\,{\rm s}^{-1}$. The fitted explosion time in this paper is MJD 60,082.25, about a half day earlier than the first detection from the amateur astronomer.
With the Monte Carlo process and involving the diffusion process, the model can fit the fast-rising evolution better, suggesting that our updated Monte Carlo method could be an efficient and important way to model the fast-rise light curves of SNe II with early flash features.

\section{Discussions and Conclusions}
\label{SecIV}

\begin{figure}[t!]
\centering
\includegraphics[width = 0.4\textwidth]{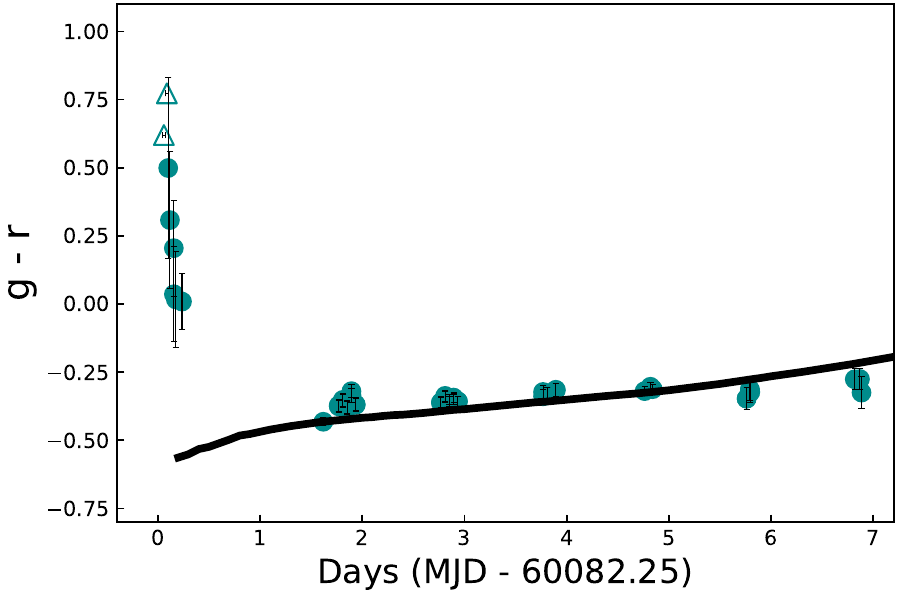}
\caption{The symbols are the $g-r$ color of SN~2023ixf from \cite{2023arXiv231114409L}, and the black line is the predicted color with our model.}
\label{fig_color}
\end{figure}

With considering the diffusion process within the unshocked CSM both for the CSM interaction and shock cooling, it seems that the early-time light curve of SN~2023ixf could reasonably be matched without the time-variant extinction adopted in \cite{2023arXiv231114409L}. However, as shown in Figure~\ref{fig_color}, our model significantly deviates from the observed $g-r$ color acquired by amateur astronomers. The predicted $g-r$ color from our model is much bluer than the observation, indicating the existence of time-variant extinction from circumstellar dust. 

\begin{figure}[htp!]
\centering
\includegraphics[width = 0.4\textwidth]{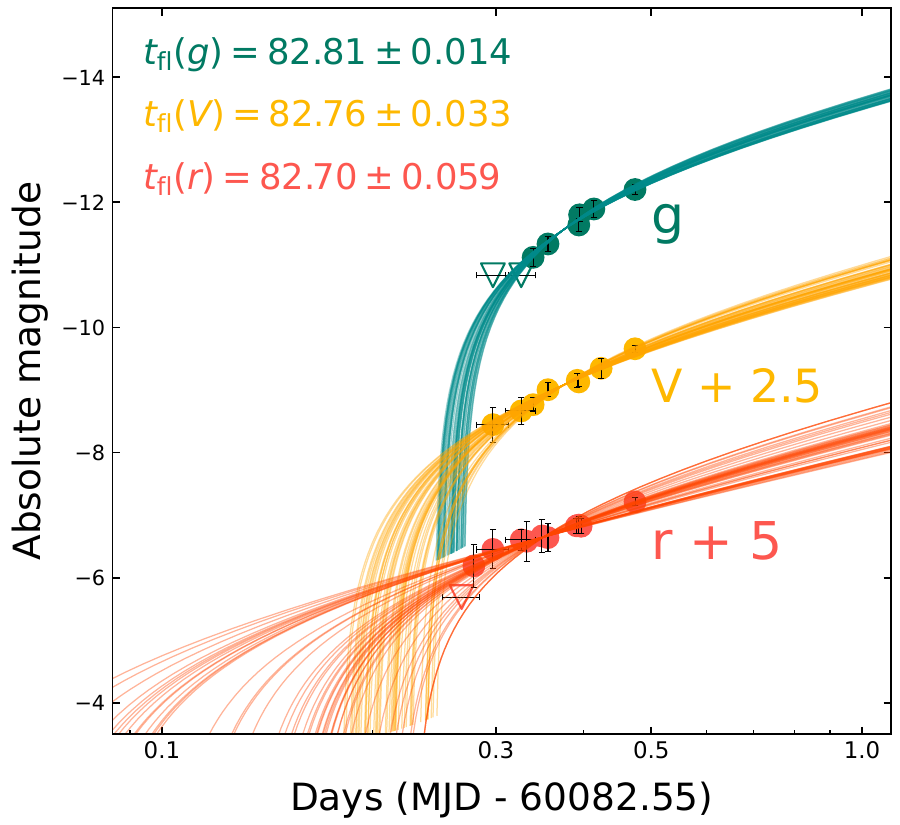}
\caption{The time of the first light ($t_{\rm fl}$) fitting for $g$ (green), $V$ (yellow), and $r$ (red) bands. The symbols are the multi-band light curve of SN~2023ixf within a few hours from the first detection by amateur astronomers. For simplicity, we adopted a linear function as $L \propto (t-t_{\rm ft})$ to estimate its single-band $t_{\rm tf}$ with the fitting region of a few hours from the first detection. The result of $t_{\rm fl}$ for each band is in a unit of days after MJD 60,000.0.  }
\label{fig_Tfl}
\end{figure}

On the other hand, in the scenario of considering the diffusion process relating to the Thomson scattering, the emergence of light is unrelated to the frequency of photons, leading to the same time of the first light for different filters. In contrast, the cross-section of the extinction and scattering relating to circumstellar dust depend on the wavelengths of photons. Therefore, the time of the first light for the red band should be earlier than that for the blue band with the existence of circumstellar dust, which is consistent with our fitting single-band $t_{\rm fl}$ as shown in Figure~\ref{fig_Tfl}.

In this paper, we have developed a Monte Carlo procedure that can simulate the photon diffusing process during the CSM interaction, in which the optical depth of unshocked CSM rapidly decreases from moderately optically thick to optically thin. This particular feature allows the photons generated at the beginning of CSM interaction to be trapped in unshocked CSM and then scattered outside of the unshocked CSM simultaneously as the photons generated by the subsequent interaction, resulting in a fast-rising light curve of SN~2023ixf soon after the explosion. The best-fit model of our CSM interaction with Monte Carlo process confirms the result in previous studies, that $\dot{M}_{\rm w}\sim 10^{-2}\ {\rm M}_{\odot}\,{\rm yr}^{-1}$ and the majority of CSM is located within $10^{15}$ cm to the SN, but with a better fit to the early-time photometric data.

One caveat in our current analysis is that we have not taken into account several factors that may affect the final light curve. In our study, the CSM is assumed to be spherically symmetric, which may deviate from the accurate distribution of the CSM surrounding SN~2023ixf as diagnosed from the spectropolarimetric observations \citep{2023ApJ...955L..37V,2024arXiv240520989S}. The conversion efficiency should be the function of CSM density with the range from 1.0 (e.g., the thick CSM as discussed in \citealt{2023arXiv230403360K}) to 0.2 (e.g., the thin CSM as discussed in \citealt{2013MNRAS.435.1520M}). Our study, however, simplified this by assuming a constant conversion efficiency, which may result in an underestimation of black-body radiation at the beginning of CSM interaction and an overestimation during subsequent interaction. Additionally, we assume the unshocked CSM is fully ionized, although the ionization of the CSM should depend on the ejecta$-$CSM interaction process and the geometric distribution of the CSM. Nevertheless, our study still provides a good understanding of CSM interaction, including converting kinetic energy into radiation and the radiative diffusion process in the unshocked CSM. This primary picture enables us to fit the early-time light curve of SN~2023ixf reasonably.

More importantly, early-phase multi-band (X-ray, Ultraviolet, optical, and radio) observations play a crucial role in uncovering the circumstellar environment and the mass-loss history of massive stars before their explosive demise. The operations of those soon-to-be-operational survey telescopes can discover a large amount of SNe with early observations (e.g., \citealt{2019ApJ...873..111I,2022Univ....9....7H,2023SCPMA..6609512W,2024ApJ...964...74S}). Therefore, our updated Monte Carlo model is significant to the upcoming multi-band light curves of SNe with confined dense CSM.


\section{Acknowledgments}

We thank Yi Yang and Gaici Li for the helpful discussions. 
This work is supported by the National Natural Science Foundation of China (NSFC grants 12403049, 12288102, and 12033003). 
Maokai Hu acknowledges the support from the Postdoctoral Fellowship Program of CPSF under Grant Number GZB20240376, and the Shuimu Tsinghua Scholar Program 2024SM118. 
X. Wang acknowledges the support from the Tencent Xplorer Prize.

%

\vspace{5mm}







\bibliography{sample631}{}
\bibliographystyle{aasjournal}



\end{document}